\documentstyle{lamuphys}
\makeatletter
\let\chapter\hid@chapter
\makeatother
\input epsf
\def \mpc       {{\rm\ Mpc}}
\def \kpc       {{\rm\ kpc}}

\def \kms       {\hbox{ km s$^{-1}$}}

\def \eg        {\hbox{\it e.g.}}

\def \etal      {{\it et al.\ }}

\def \Ho        {{\rm\ H_{o}}}
\def \kmsmpc    {{\rm\ km\ s^{-1}\ Mpc^{-1}}}

\def \msol      {{\rm M}_\odot}

\def \hinv      {\hbox{$\, h^{-1}$} }

\def \se        {\!=\!}

\newcount\itemno
\def \ino         { \the\itemno\global\advance\itemno by 1 }
\def\apj{ApJ}

\def\mn{MNRAS}

\begin{document}
\pagenumbering{arabic}
\title{The Tully--Fisher Relation : A Numerical Simulator's Perspective}

\author{August E. Evrard\inst{1,2,3}}
\institute{Physics Department, University of Michigan, 48109-1120 USA 
\and 
Institut d'Astrophysique, 98bis Blvd Arago, 75014 Paris, France
\and 
Max--Planck--Institut f\"ur Astrophysik, Karl--Schwarz.--Str. 1,
Garching bei M\"unchen, Germany
}

\maketitle
\begin{abstract}
In this brief contribution, I will outline hopes of understanding the
origin of galaxy scaling relations using numerical simulations as a
tool to gain understanding.  The case of the Tully--Fisher relation
for disk galaxies is used as a working example to illustrate the 
modest achievements to date and the difficult tasks ahead. 
\end{abstract}
\section{Overview}

Galaxies, like people, come in a variety of shapes and sizes.  
Like people, galaxies tend to have many features in common, but
explaining in detail the origin of these features can be a 
difficult task.  Understanding why the luminosity of a disk galaxy 
should scale as a power of its circular velocity is a bit like
understanding the relation between a person's weight and his or her
belt length.  There are basic governing principles --- bodies are
(crudely speaking) similarly shaped bags of water and galaxies are
(crudely speaking) similarly structured gravitational objects ---
but other factors creep in when you start giving it more 
careful thought.  A skinny basketball player and a jockey might 
have the same waistline, but their body masses could differ by more
than $50\%$ because of their height difference.  Why couldn't 
two galaxies lying in potential wells of the  same circular speed 
differ in luminosity by, say, a factor 3 because of differences in 
their gas dynamic/stellar evolutionary histories?   It appears nature
does not allow this to happen; the scatter about the Tully--Fisher
relation is remarkably small (see Giovanelli and others in this
proceedings).

The remarkable nature of such a tight correlation in non--trivially
linked physical properties is made apparent when one considers their
complicated birthing process shown schematically in Figure~1.  
This picture was laid out theoretically in the late 1970's, and the 
seminal paper of White and Rees (1978) cemented the elements together
within a modern, hierarchically clustering framework.  In hierarchical
models, gravity amplifies density perturbations on ever--increasing
mass and length scales, driving an overdense, filamentary/knotty  
network which evolves in a nearly self--similar fashion.  On mass
scales roughly between $10^8$ and $10^{12} \msol$ and in the absence
of significant non--gravitational heat input, cooling via radiative
processes removes thermal pressure support from the baryons.  This
process acts to concentrate the baryons within an assumed, dominant
halo of dark matter.  Once self--gravitating, star formation is
ignited in a manner poorly understood from first principles (hence the
``black box'' in the figure) and {\sl poof!\/} we end up with a disk 
galaxy rotating in its dominant dark halo.  

Given that rotation speed is a direct measure of total mass within a
fixed density contrast $M_\delta$ (see below), then a suspiciously simple
interpretation of the tight, observed Tully--Fisher relation in the
context of Figure~1 is that cooling and star formation are highly 
regular and dependent primarily on $M_\delta$.  Such a simple picture
appears to be true for the structure of collisionless halos formed
from hierarchical, gravitational clustering.  A single characteristic
function with parameters smoothly varying with mass appears to
describe the density and velocity structure of collapsed objects 
(see White's contribution in this volume).  

The situation in Figure~1 is simplistic in a number of ways.  
Formation of a single galaxy actually entails a network of such
segments, inter-connected in a manner reflecting the particular 
merger history of that object.  An ensemble of equal mass objects
observed today will naturally arise from a variety of merger 
histories/inter-connections.  Why doesn't this variety evidence itself
as a large scatter in the Tully--Fisher relation? 
(Eisenstein in this volume presents a similar argument from a slightly
different perspective.)

\begin{figure}
\vskip -1.0 truecm
\epsfxsize=12.0cm
\epsfysize=12.0cm
\vbox {\hskip 1.0truecm \epsfbox{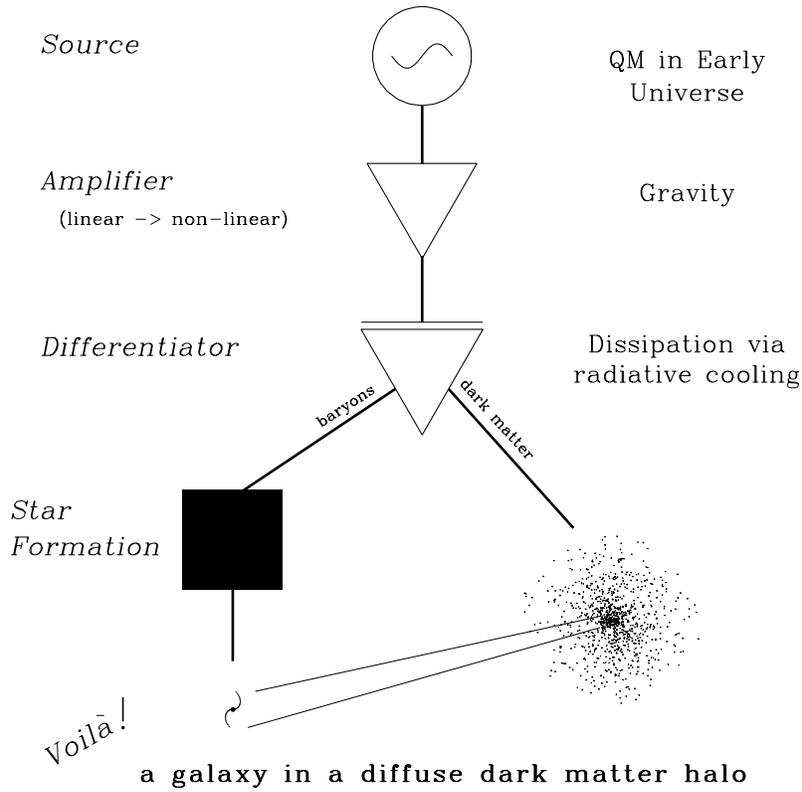} } 
\vskip -0.5 truecm
\caption{
Schematic illustrating the basic principles behind the White--Rees
picture of galaxy formation discussed in the text.  
}

\end{figure}

\section{Looking Behind}

Numerical simulations with gas dynamics are now beginning to be 
used to address the origin of disk galaxies and 
the Tully--Fisher relation (Katz \& Gunn 1991; Evrard,
Summers \& Davis 1994, Navarro \& White 1994, Steinmetz \& M\"uller
1994; Navarro \& Steinmetz 1996;
Tissera, Lambas \& Abadi 1996; Groom 1997).  Most of these simulations
ignore star formation altogether.  Those in which it was
included failed to form a disk of stars, forming spheroids instead.  
So the best we can do at the moment is analyze the gas disk
properties.  An idea of where we stand
is shown in Figure~2, which compares the cold, gas mass in the galaxy 
to its circular speed.  Data shown are from Navarro \& White (1994; 
hereafter NW) and
from a unpublished P3MSPH simulation by myself of a random, $16 \mpc$ 
($\Ho \se 50 \kmsmpc$) patch of a standard cold dark matter universe.
The characteristics of the simulation are identical to that detailed
in Evrard \etal (1994), with the exception of it being a random,
(instead of constrained cluster) spatial region and it being evolved
to the present (instead of $z \se 1$).  The interested reader should
consult these papers for further details.   The ``raw'' points in the
figure use the peak in the measured circular speed of the gas disk,
while the ``corrected'' points enforce centrifugal equilibrium at that
point in the rotation curve. The correction is necessary because the
size of the disks is within a factor of a few of the spatial 
resolution limit of the simulation.  The dotted line in the figures
has slope $2.45$.  

\begin{figure}
\vskip -4.4 truecm
\epsfxsize=12.0cm
\epsfysize=14.0cm
\vbox{ \hskip -0.1 truecm \epsfbox{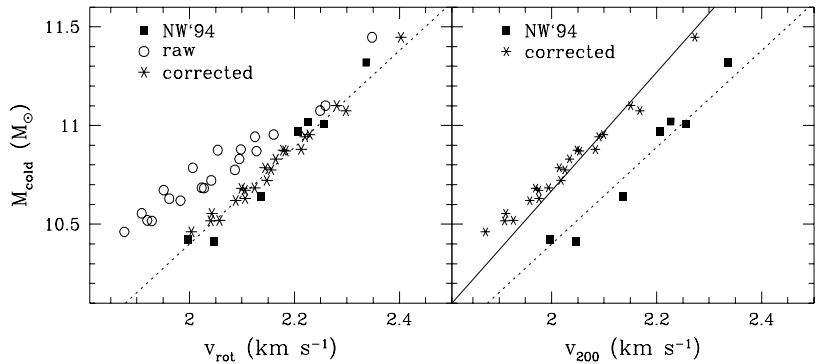}}
\vskip -5.0 truecm
\caption{
Cold gas mass versus rotation speed derived from gas dynamic
simulations of Navarro \& White 1994 (NW) and Evrard (1995,
unpublished).  See text for a discussion.  
}
\end{figure}

The good news from Figure~2 is that the scatter in both data sets 
is quite small, 
in fact, smaller by a factor 2 than typical observed values.  The bad
news is that the left hand panel of the figure is a dishonest
comparison of two independent experiments.  The NW rotation 
speed is actually $\sqrt{GM/r}$ measured at a density contrast of 200
with respect to the critical background.  Going back and measuring
the same quantity for the disk galaxies in the P3MSPH simulation 
(the ``corrected'' values in the right panel) results in a 
systematic offset in the intercept of the two data sets.  This
intercept is not due to different values of Hubble constant, both use 
$h\se 0.5$, or cosmic baryon fraction, since both assume $\Omega_b \se 0.1$ 
and $\Omega \se 1$.  In this cosmology, the radius and total 
mass defining a density contrast of 200 at the present epoch are 
\begin{equation} 
r_{200} \ = \sqrt{\frac{2}{\delta}} \ \frac{v_{200}}{\Ho} \ = \ 100 \ 
 \biggl(\frac{v_{200}}{100 \kms} \biggr)  \hinv \kpc 
\end{equation}
\begin{equation} 
M_{200} \ = \ 2.32 \times 10^{11} \ \biggl(\frac{v_{200}}{100 \kms} \biggr)^3 
\hinv \msol 
\end{equation}
If all the baryons within this density contrast cool and sink to the
center of the halo, then the cold, galactic mass will simply be
$\Omega_b M_{200}$.  This implies a Tully--Fisher relation shown as
the solid line in the right panel (using $h \se 0.5$ as in the
simulations).  One interpretation of this panel
is that the P3MSPH treatment is allowing nearly all the gas to cool in
the halos while NW's treatment allows half the baryons to cool, with
the remainder in a tenous, hot halo.  It remains to be seen if this
interpretation is correct but, at any rate, the offset between the
two data sets is most likely numerical in origin, since both are
attempting to model essentially identical physical situations.  
The silver lining here is that the small degree of scatter in the
relation appears insensitive to the detailed numerical treatment.  

\section{Looking Ahead}

The example above illustrates our current level of uncertainty in
modeling just some of the physical processes associated with Figure~1. 
The black box of star formation is largely unexplored territory.
Presumably different physical and numerical parameterizations for star
formation and feedback will lead to an even larger range of possible 
answers than that illustrated in Figure~2. 

On the bright side, a comparison between codes attempting to model 
the branch above the differentiator in Figure~1 (gravitational 
clustering without radiative cooling) indicate 
there is quite good agreement in the gas and dark matter solutions
over the dynamic range presently accessed by such experiments, 
that is, density contrasts up to about $10^4$ (Frenk, White \etal 1997, in
preparation).  Similar comparisons including cooling will ultimately 
enable sorting out of physical versus numerical effects.  

In the realm of galaxy scaling relations, theorists are in the 
typical position of attempting to understand current 
observations; predictive power is tenuous at best.  
From the excellent new data presented at this meeting, 
particularly in the area of evolution in the scaling relations at 
moderate to high redshift (\eg, the contributions of Franx, Pahre, Schade,
Guzman, Dickinson, and Ziegler in this volume), 
it seems the observers are accelerating their pace!  
Modeling this wealth of data presents a formidable challenge for the
foreseeable future.

\bigskip
\bigskip

This work was supported in the USA by NASA through grant 
NAG5--2790 and in France by the CIES and CNRS at the 
Institut d'Astrophysique in Paris.  
It is a pleasure to thank L. DaCosta, A. Renzini and the rest of the
organizers for putting together an excellent workshop.  
I am most grateful to S. White for hospitality extended during my stay
at MPA, where this proceedings was written.


\begin{thebibliography}

\bibitem { } { } { } 
Evrard, A.E., Summers, F.J. \& Davis, M. 1994, \apj, 422, 11

\bibitem { } { } { } Groom, W. 1997, PhD Thesis, Cambridge University.

\bibitem { } { } { } Katz, N. \& Gunn, J.E. 1991, \apj, 377, 365

\bibitem { } { } { } Navarro, J. \& White, S.D.M. 1996, \mn, 267, 401

\bibitem { } { } { } Navarro, J. \& Steinmetz, M. 1996, \apj submitted 
(astro-ph/9605043)

\bibitem { } { } { } Steinmetz, M. \& M\"uller, E. 1994, AA, 268, 391 

\bibitem { } { } { } Tissera, P.B., Lambas, D.G, Abadi, M.G. 1996,
\mn, submitted (astro-ph/9603153)

\bibitem { } { } { } White, S.D.M. \& Rees, M. 1978, \mn, 183, 341

\end{thebibliography}
\end{document}